\documentclass[11pt,letterpaper]{article}

\usepackage{latexsym}
\usepackage{amssymb,amsmath, bm,pgfplots,tikz,bbm}
\usepackage{graphicx}
\usepackage{algpseudocode}
\usepackage{marvosym}
\usepackage{multirow,float}
\usepackage{subcaption}
\usepackage{comment}
\usepackage{centernot}
\usepackage{makecell}
\usepackage{natbib}
\usepackage{color}
\usepackage[bookmarksopen=true, bookmarksnumbered=true,
pdfstartview=FitH, breaklinks=true, urlbordercolor={0 1 0}, citebordercolor={0 0 1}]{hyperref}

\usepackage{dcolumn}
\newcolumntype{.}{D{.}{.}{-1}}
\newcolumntype{d}[1]{D{.}{.}{#1}}
\newcolumntype{C}{>{$}c<{$}}

\usepackage{theorem}
\theoremstyle{plain}
\theoremheaderfont{\scshape}

\newcommand{\qed}{\hfill \ensuremath{\Box}}

\usetikzlibrary{decorations.markings}
\usetikzlibrary{decorations.pathmorphing}
\usetikzlibrary{shapes.geometric, arrows}
\usetikzlibrary{arrows,decorations.pathmorphing,backgrounds,positioning,fit,matrix}
\usetikzlibrary{shapes,decorations,arrows,calc,arrows.meta,fit,positioning}
\tikzset{auto,node distance =1 cm and 1 cm,semithick,
	state/.style ={circle, draw, minimum width = 0.7 cm},
	point/.style = {circle, draw, inner sep=0.04cm,fill,node contents={}},
	bidirected/.style={Latex-Latex,dashed},
	el/.style = {inner sep=2pt, align=left, sloped}
}
\usetikzlibrary{positioning}
\usetikzlibrary{fadings}
\usetikzlibrary{intersections}
\usepackage{kantlipsum}
\allowdisplaybreaks

\usepackage{rotating}

\usepackage{arydshln}
\usepackage{threeparttable}

\usepackage[compact]{titlesec}

\newcommand{\blind}{0}

\usepackage{algorithm}
\usepackage{algpseudocode}
\usepackage{pifont}
\usepackage{xcolor}

\begin{document}

\newcommand\ud{\mathrm{d}}
\newcommand\dist{\buildrel\rm d\over\sim}
\newcommand\ind{\stackrel{\rm indep.}{\sim}}
\newcommand\iid{\stackrel{\rm i.i.d.}{\sim}}
\newcommand\logit{{\rm logit}}
\renewcommand\r{\right}
\renewcommand\l{\left}
\newcommand\cO{\mathcal{O}}
\newcommand\cY{\mathcal{Y}}
\newcommand\cZ{\mathcal{Z}}
\newcommand\E{\mathbb{E}}
\newcommand\cL{\mathcal{L}}
\newcommand\V{\mathbb{V}}
\newcommand\cA{\mathcal{A}}
\newcommand\cB{\mathcal{B}}
\newcommand\cD{\mathcal{D}}
\newcommand\cE{\mathcal{E}}
\newcommand\cM{\mathcal{M}}
\newcommand\cU{\mathcal{U}}
\newcommand\cN{\mathcal{N}}
\newcommand\cT{\mathcal{T}}
\newcommand\cX{\mathcal{X}}
\newcommand\bA{\mathbf{A}}
\newcommand\bH{\mathbf{H}}
\newcommand\bB{\mathbf{B}}
\newcommand\bP{\mathbf{P}}
\newcommand\bQ{\mathbf{Q}}
\newcommand\bU{\mathbf{U}}
\newcommand\bD{\mathbf{D}}
\newcommand\bS{\mathbf{S}}
\newcommand\bx{\mathbf{x}}
\newcommand\bX{\mathbf{X}}
\newcommand\bV{\mathbf{V}}
\newcommand\bW{\mathbf{W}}
\newcommand\bM{\mathbf{M}}
\newcommand\bZ{\mathbf{Z}}
\newcommand\bY{\mathbf{Y}}
\newcommand\bt{\mathbf{t}}
\newcommand\bbeta{\bm{\beta}}
\newcommand\bpi{\bm{\pi}}
\newcommand\bdelta{\bm{\delta}}
\newcommand\bgamma{\bm{\gamma}}
\newcommand\balpha{\bm{\alpha}}
\newcommand\bone{\mathbf{1}}
\newcommand\bzero{\mathbf{0}}
\newcommand\tomega{\tilde\omega}
\newcommand{\argmax}{\operatornamewithlimits{argmax}}

\newcommand{\R}{\textsf{R}}

\newcommand\spacingset[1]{\renewcommand{\baselinestretch}%
{#1}\small\normalsize}

\spacingset{1}

\newcommand{\tit}{\bf Comment on ``Generic machine learning inference on heterogeneous treatment effects in randomized experiments.''}


\if1\blind
\title{\tit}
\fi

\if0\blind

{\title{\tit}

  \author{
  Kosuke Imai\thanks{Professor, Department of Government and
      Department of Statistics, Harvard University, Cambridge, MA
      02138. Phone: 617--384--6778, Email:
      \href{mailto:Imai@Harvard.Edu}{Imai@Harvard.Edu}, URL:
      \href{https://imai.fas.harvard.edu}{https://imai.fas.harvard.edu}} \hspace{.75in}
Michael Lingzhi Li\thanks{Assistant Professor, Technology and Operations Management, Harvard Business School, Boston, MA
	02163. \href{mailto:mili@hbs.edu}{mili@hbs.edu}}
}

  \date{\today
  }

  \fi
\maketitle

\pdfbookmark[1]{Title Page}{Title Page}

We congratulate the authors --- Victor Chernozhukov, Mert Demirer,
Esther Duflo, and Iv\'{a}n Fern\'{a}ndez-Val (hereafter CDDF) --- for
their excellent contribution to the growing literature on the
estimation of heterogeneous treatment effects.  CDDF's proposed
approach is both methodologically innovative and practically relevant.
Their method also comes with three important advantages.  First, it
can incorporate essentially any machine learning (ML) model.  Second,
the proposed methodology does not require that the ML model accurately
estimates the conditional average treatment effect (CATE).  Finally,
it quantifies uncertainty including the one that arises from data
splitting used to avoid overfitting the ML model.  We believe that
these attractive features will enable applied researchers to
effectively but safely explore heterogeneous treatment effects in
randomized experiments while leveraging the power of modern ML
algorithms.

In this commentary, we focus on CDDF's key contribution---the
development of uncertainty quantification methodology.  The authors
correctly point out that while data splitting is a standard technique
to avoid overfitting ML models, the resulting causal estimates can
vary substantially across different random splits of the data.  In
practice, this leads to an undesirable consequence: one may obtain
different empirical conclusions, depending on the random seed.  This
practical problem motivates the need to quantify the uncertainty that
arises from data splitting as well as the estimation uncertainty
conditional on the data split.

As a solution to this problem, CDDF proposes the split-sample robust
inference (SSRI) methodology that requires randomly splitting the data
many times.  Then, for each split, researchers obtain an estimated
causal quantity of interest and its conditional confidence interval.
The authors show that a valid unconditional confidence interval, which
accounts for the randomness of data splitting, can be constructed by
simply using the median of lower conditional confidence bounds and
that of upper conditional confidence bounds.

\paragraph{An alternative randomization inference approach.}
One potential limitation empirical researchers may face when applying
the SSRI method in practice is that repeated data splitting increases
computational burden.  Since fitting a complex ML model to even a
single data split can be computationally expensive, repeating this
process many times is not desirable.  Moreover, ML models often require parameter tuning, which in
itself may require another data splitting and incur additional
computational cost.  Given CDDF's recommendation that researchers
split the data 250 times and fit a ML model to each split, the SSRI
methodology can be, in practice, quite computationally intensive.

In a separate paper, we developed an alternative, randomization
inference (RI) methodology that addresses this remaining challenge
\citep{imai:li:22}.  Like the SSRI method, the RI methodology is
applicable to a generic ML model regardless of whether it estimates
the CATE well.  Indeed, this alternative approach essentially requires
only three randomization assumptions---random assignment of treatment,
random sampling of units, and random splitting of data.  This is
consistent with the design-based approach pioneered by \cite{neym:23}.
Unlike the SSRI method, however, the RI approach does not require
repeated data splitting and hence is computationally more efficient.

Here, we briefly discuss the RI methodology while referring the
readers to \citet{imai:li:22} for details.  For simplicity, suppose
that we have $K$ sorted groups of equal size and the data are randomly
divided into $L \ge 2$ splits of equal size.  To both avoid
overfitting and gain statistical efficiency, the RI method uses the
standard cross-fitting procedure.  That is, every split, denoted by
$\mathcal{L}_\ell$, is used once as the evaluation data to estimate
the GATES $\hat\gamma_k^{(\ell)}$, while the remaining $L-1$ splits,
denoted by $\mathcal{L}_{-\ell}$, are used to estimate the ML proxy
$S^{(-\ell)}$.  This process is repeated for each split, and the final
estimate is given by the following simple average across all splits,
i.e.,
\begin{equation}
  \hat\gamma \ = \ \frac{1}{L} \sum_{\ell=1}^L \hat\gamma_k^{(\ell)}. \label{eq:gamma_hat}
\end{equation}

Using CDDF's notation whenever possible, a natural estimator of the
GATE for the $k$th group based on the $\ell$th split (evaluation data)
is given by,
\begin{equation}
  \hat\gamma_k^{(\ell)} \ = \ \frac{K}{N_1^{(\ell)}} \sum_{i \in \mathcal{L}_\ell} Y_i D_i \hat{f}_k^{(-\ell)}(Z_i)
  - \frac{K}{N_0^{(\ell)}} \sum_{i\in \mathcal{L}_\ell}
  Y_i(1-D_i)\hat{f}_k^{(-\ell)}(Z_i) \label{eq:gammak_hat}
\end{equation}
where
$N_d^{(\ell)} = \sum_{i \in \mathcal{L}_\ell} \mathbf{1}\{D_i = d\}$
for $d=0,1$ denotes the treatment/control group size in the evaluation
data,
$\hat{f}_k^{(-\ell)}(Z_i) \ = \ \mathbf{1}\{S^{(-\ell)}(Z_i) \geq
\hat{c}_{k}^{(\ell)} \} - \mathbf{1}\{S^{(-\ell)}(Z_i) \geq
\hat{c}_{k-1}^{(\ell)} \}$ is the sorted group indicator variable
based on the ML proxy obtained from the remaining splits, and
$\hat{c}_{k}^{(\ell)} \ = \ \inf \{c \in \mathbb{R}: \sum_{i \in
  \mathcal{L}_\ell} \mathbf{1}\{S^{(-\ell)}(Z_i)>c\} \le
|L_\ell|k/K\}$ is the estimated $k$th quantile for the evaluation
data.

Equation~\eqref{eq:gammak_hat} resembles the standard estimator for
evaluating the expected utility of estimated individualized treatment
rules (ITR) represented by $\hat{f}_k^{(-\ell)}(Z_i)$.  This relation
enables us to leverage the result from the experimental evaluation of
ITR \citep{imai:li:23}.  In particular, Theorem~5 of
\citet{imai:li:22} presents the finite sample bias and variance of the
estimator given in Equation~\eqref{eq:gamma_hat} for the following
estimand, which is a variant of the GATES,
\begin{equation}
  \gamma_k \ = \ \E\{\E(Y_i(1)-Y_i(0) \mid \hat{c}_k \le S(Z_i) \le \hat{c}_{k+1})\}.
\end{equation}
In this equation, the inner expectation is taken over the distribution
of $\{Y_i(1), Y_i(0), Z_i\}$ that fall within the sorted group based
on a training data set while the outer expectation is taken over the
random sampling of training data.  Based on this result, one can
construct the asymptotic confidence interval.  Importantly, while the
resulting confidence interval accounts for both estimation and data
splitting uncertainties, it only requires a single round of data
splitting into $L$ splits.

To derive the variance expression, we must account for the fact that
the cross-fitting procedure uses the same data to both train the ML
algorithm and estimate the GATES.  We apply the useful result of
\citet[][Lemma~1]{nadeau2000inference}, which implies the following,
\begin{equation}
  \V(\hat\gamma_k) \ = \ \V(\hat\gamma_k^{(\ell)}) -
  \frac{L-1}{L}\E(V^2_{k}), \label{eq:var}
\end{equation}
for each sorted group $k$ where
$V^2_k = \sum_{\ell=1}^L (\hat\gamma_k^ {(\ell)} -
\hat\gamma_k)^2/(L-1)$ is the sample variance of split-specific GATE
estimate.  While the first term of Equation~\eqref{eq:var} denotes the
variance of GATE estimate based on a single split, the second term
represents the efficiency gain due to cross-fitting. Thus, the
efficiency gain due to cross-fitting is large when the split-specific GATE estimate has a
large variance across splits.

Finally, we note that CDDF's classification analysis of CLAN can also
be conducted using this estimator by replacing the outcome $Y_i$ with
a quantity of interest $g(Y_i, Z_i)$.  CLAN compares the two groups
that are predicted to benefit most and least from the treatment, i.e.,
$\gamma_K$ and $\gamma_1$.  This means that we need to derive the
covariance between $\hat\gamma_K$ and $\hat\gamma_1$.  Theorem~7 of
\cite{imai:li:22} derives this covariance as a part of the
nonparametric test of treatment effect heterogeneity.

\paragraph{Comparing the finite-sample performance through a
  simulation study.}
We compare the finite-sample performance of the SSRI confidence
interval with that of the RI confidence interval. We compute the
latter using the open-source software package evalITR \citep{evalITR}.
Unlike the simulation study presented in CDDF, which assumes a linear
interactive model, we randomly selected three data generating
processes out of the 77 settings used for the 2016 Atlantic Causal
Inference Conference Competition \citep{dori:etal:19}, including one
linear, polynomial, and step-change treatment effect model (No. 13,
28, and 66).  These three simulation settings cover various levels of
treatment effect heterogeneity and different functional forms.  We use
the empirical distribution of covariates in the full sample ($n=4802$)
as their population distribution. See \cite{imai:li:22} for additional
details about the simulation setup.

We focus on the properties of confidence intervals for GATES with
$K=5$ groups using a total sample size of 100, 500, and 2500. For all
simulation experiments, we train LASSO to estimate the CATE.  We
choose LASSO here because other more complex ML algorithms such as
BART and Causal Forest are computationally too expensive for repeated
sample splitting required by the SSRI method. For the SSRI method, we
follow CDDF's empirical application and utilize a $67\%$/$33\%$ split
and set the number of repeated data splitting to 250.  The results
presented below are based on 1,000 simulated data sets (the true
values are approximated by averaging over 10,000 simulated data sets).

To compare the SSRI method with the RI method on an equal footing, we
present the results based on the mean metric.  While CDDF recommend
the use of the median metric, they also note that the mean metric
enjoys the same properties as the median (page~22).  Indeed, we find
that the results based on the median metric are essentially
identical. Following the advice of CDDF (page~25), we further test a
$80\%$/$20\%$ split so that the auxiliary data set is much larger than
the main data set, making it more likely to satisfy one of the SSRI's
key assumptions (R3). Finally, for the sake of direct comparison, we
set the number of splits to $L=3$ and $L=5$ for the RI method,
implying that the split between training and evaluation is
$67\%$/$33\%$ and $80\%$/$20\%$ respectively. In both cases, the
SSRI method was approximately $40$ times more computationally
intensive than the RI method. 

\begin{figure}[t!]
    \centering 
    \includegraphics[width=\textwidth, trim = 0 40 0 0, clip]{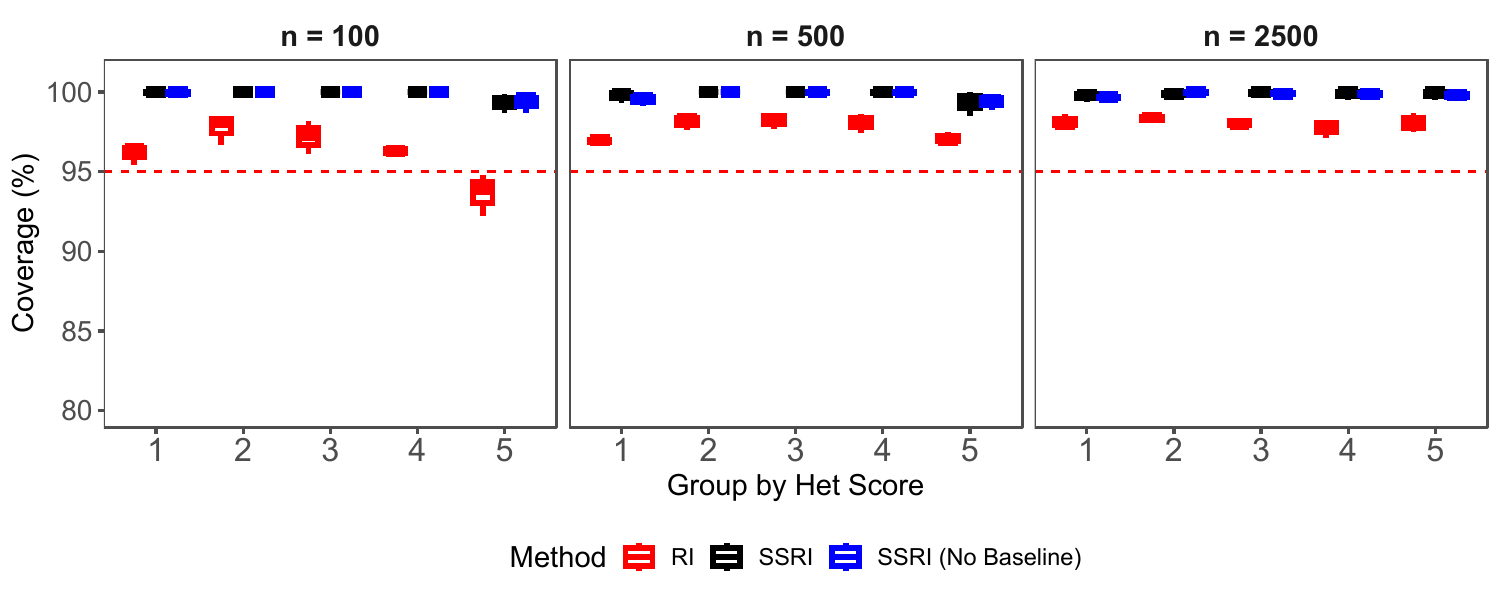}
    \includegraphics[width=\textwidth]{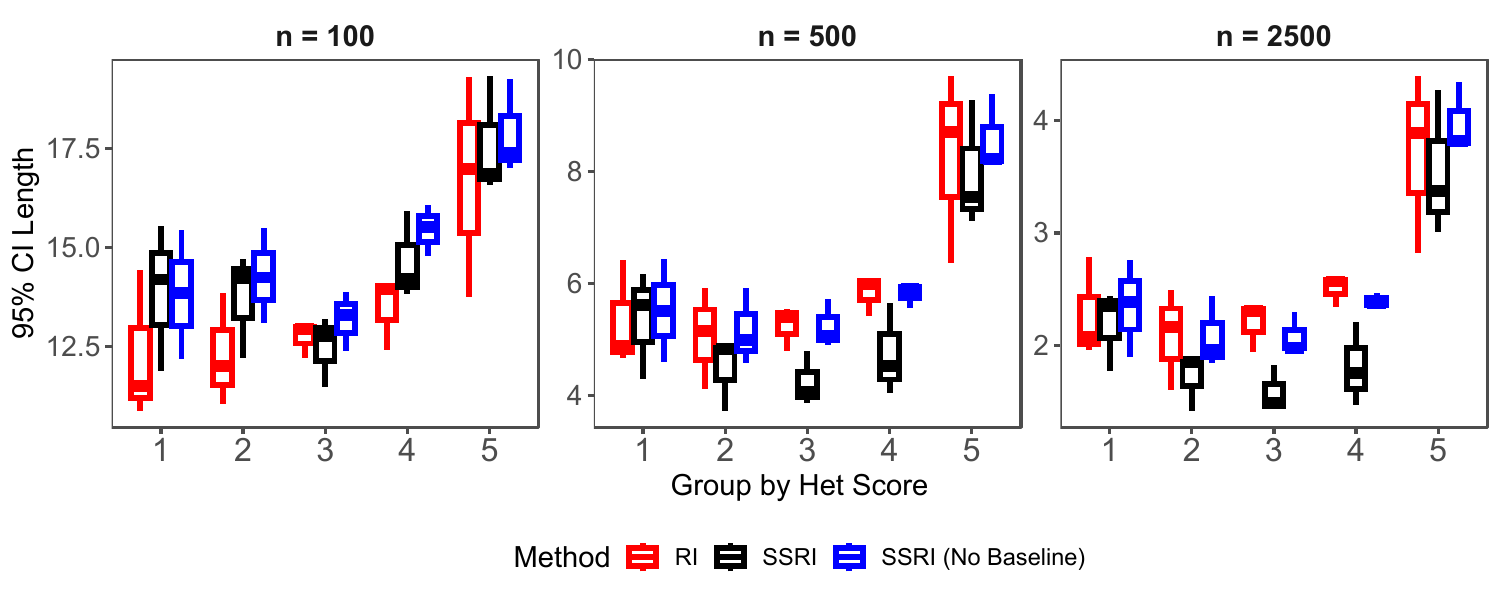}
    \caption{Empirical coverage (upper panel) and average length
      (lower panel) of 95\% confidence intervals for CDDF's
      Split-Sample Robust Inference (SSRI) confidence intervals
      (black) and our Randomization-Inference (RI) confidence
      intervals (red) under the $67\%$/$33\%$ split scenario.  For
      direct comparison with the RI confidence intervals, we also
      include the SSRI confidence intervals without baseline
      estimation (blue).}
		\label{fg:multi_3}
\end{figure}

\begin{figure}[t!]
  \centering
  \includegraphics[width=\textwidth, trim = 0 40 0 0, clip]{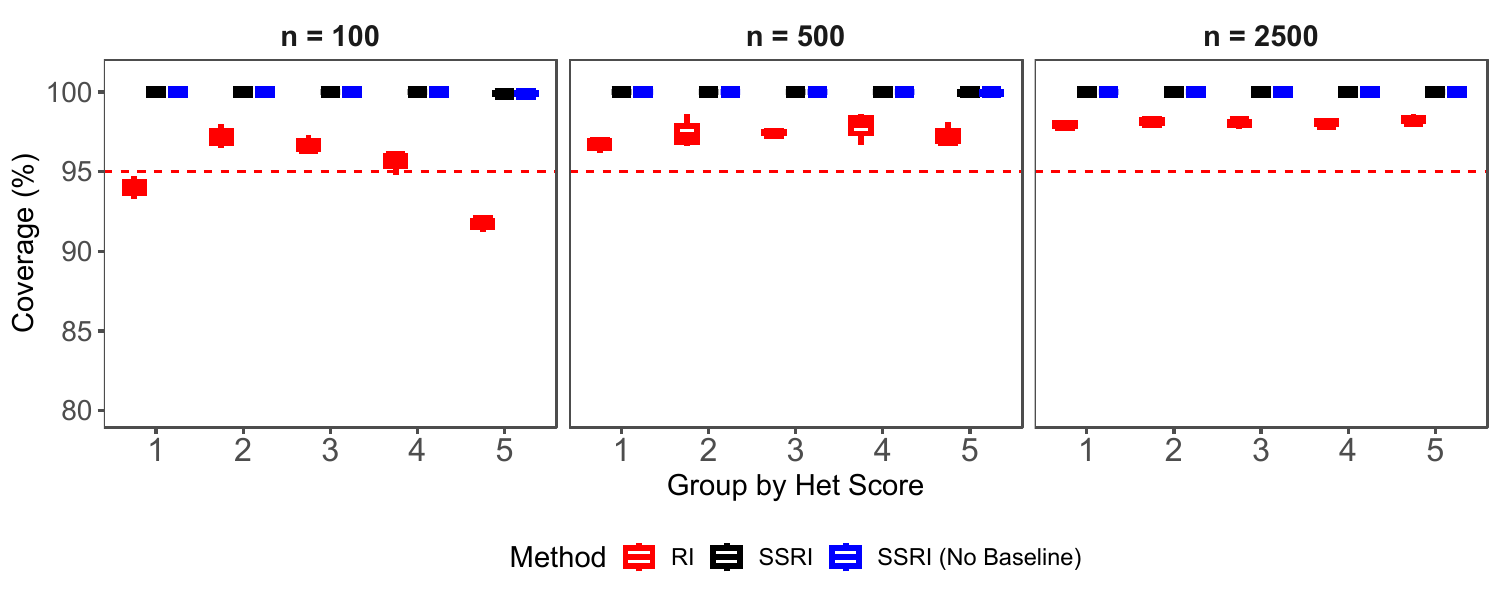}
  \includegraphics[width=\textwidth]{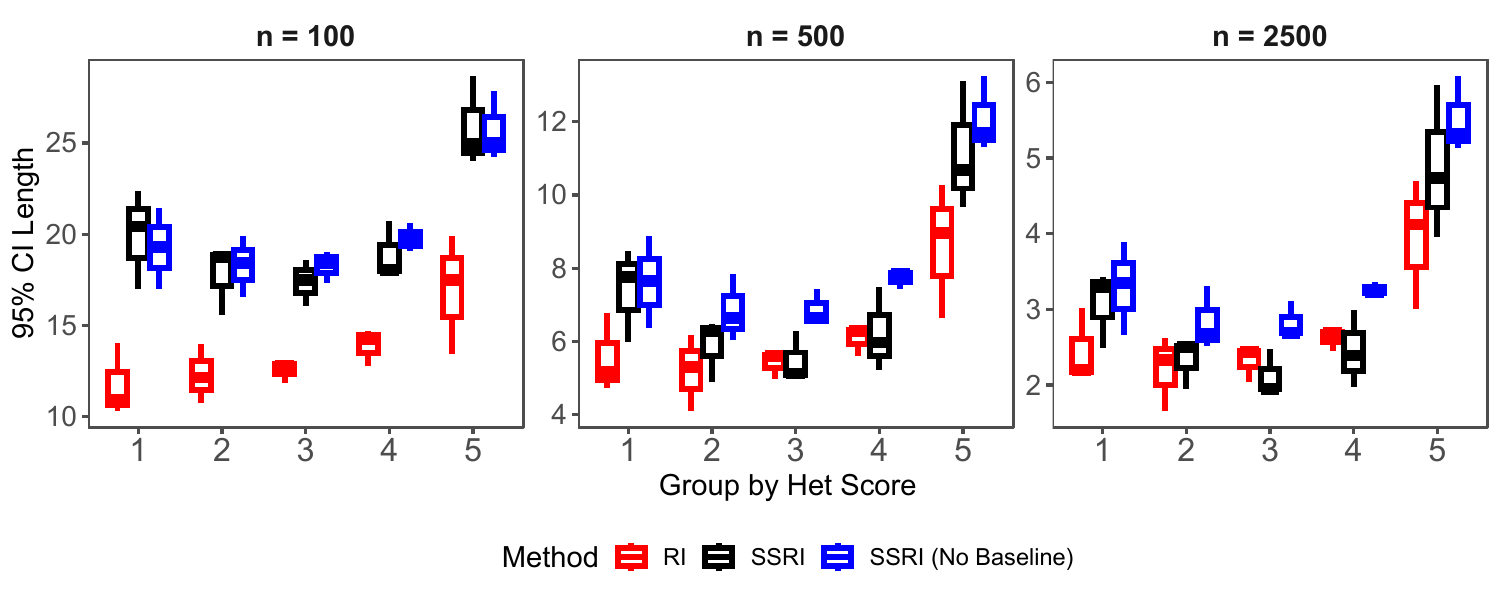}
  \caption{Empirical coverage (upper panel) and average length (lower
    panel) of 95\% confidence intervals for CDDF's Split-Sample Robust
    Inference (SSRI) confidence intervals (black) and our
    Randomization-Inference (RI) confidence intervals (red) under the
    $80\%$/$20\%$ split scenario.  For direct comparison with the RI
    confidence intervals, we also include the SSRI confidence
    intervals without baseline estimation (blue).}
		\label{fg:multi_5}
\end{figure}


Figures~\ref{fg:multi_3}~and~\ref{fg:multi_5} present the results for
$67\%/33\%$ and $80\%/20\%$ splits, respectively.  For direct
comparison with RI, we use the mean point estimate of SSRI, rather
than the estimate based on median as recommended by CCDF. The upper
panels of these figures present the empirical coverage of 95\%
confidence intervals based on the RI (red) and SSRI (black)
methods. Given that the SSRI method utilizes another ML model for
estimating the baseline conditional average function, we also examine
the performance of SSRI without baseline estimation (blue) for direct
comparison with RI.  We find that the SSRI confidence interval (with
or without baseline estimation) is conservative across sorted groups
in all sample sizes, leading to nearly 100\% coverage in every case.

Although it is difficult to pinpoint the exact source of this
conservativeness, we hypothesize that it is due to the use of Markov's
inequality in CCDF's theoretical analysis. While the inequality is
tight for a general random variable, it may be conservative in
realistic data generating settings including those of the ACIC data
competition setup used in our paper. In contrast, while the RI
confidence interval also tends to be conservative, it exhibits a
smaller degree of overcoverage than the SSRI confidence interval.

The conservativeness of the SSRI method leads to wider confidence
intervals.  The bottom panels of
Figures~\ref{fg:multi_3}~and~\ref{fg:multi_5} show that the average
length of the RI confidence interval is generally shorter than the
SSRI method when no extra ML baseline estimation is used. In theory,
the extra information about the estimator distribution obtained from
SSRI's repeated sampling procedure should help construct a shorter
confidence interval than RI.  In other words, RI's computational gain
may come at the expense of confidence interval length.  In practice,
however, the conservativeness of the estimator appears to largely
erase this benefit in our simulations.

The baseline estimation helps shrink the SSRI confidence intervals,
suggesting that a similar strategy might also improve the RI
confidence intervals. For a large sample size, the average length of
the SSRI confidence intervals with baseline estimation is roughly
comparable to that of the RI confidence intervals without.  Finally,
increasing the relative size of training data does not significantly
affect the RI confidence intervals but greatly widens the SSRI
confidence intervals.
      
\paragraph{Concluding remarks.}  We commend CDDF for proposing novel
ways to explore heterogeneous effects with new causal quantities such
as GATES and CLAN while developing a statistical method to quantify
estimation uncertainty.  We expect their work to have significant
impacts on both applied empirical research and methodological
literature.  In this commentary, we discussed an alternative,
randomization-inference approach to the same problem.  We hope that
this alternative perspective inspires further methodological
development.  Finally, many interesting open problems remain.  For
example, future work should consider the data-driven selection of
cutoffs used to define sorted groups as well as the exploration of
heterogeneous treatment effects in adaptive experiments.

\newpage
\pdfbookmark[1]{References}{References}
\bibliography{sample,my,imai}

\end{document}